\documentclass[twocolumn,aps,prl]{revtex4}
\usepackage{amsmath}
\usepackage{amsfonts}
\usepackage{amssymb}
\usepackage{amsthm}
\usepackage{graphicx}

\begin{document}
\title{The theory of magnetic field induced domain-wall 
propagation in magnetic nanowires}
\author{X. R. Wang}
\affiliation{Physics Department, The Hong Kong University of
Science and Technology, Clear Water Bay, Hong Kong SAR, China}
\author{P. Yan}
\affiliation{Physics Department, The Hong Kong University of
Science and Technology, Clear Water Bay, Hong Kong SAR, China}
\author{J. Lu}
\affiliation{Physics Department, The Hong Kong University of
Science and Technology, Clear Water Bay, Hong Kong SAR, China}
\author{C. He}
\affiliation{Physics Department, The Hong Kong University of
Science and Technology, Clear Water Bay, Hong Kong SAR, China}
\begin{abstract}
A global picture of magnetic domain wall (DW) propagation 
in a nanowire driven by a magnetic field is obtained: 
A static DW cannot exist in a homogeneous magnetic 
nanowire when an external magnetic field is applied. 
Thus, a DW must vary with time under a static magnetic field. 
A moving DW must dissipate energy due to the Gilbert damping. 
As a result, the wire has to release its Zeeman energy 
through the DW propagation along the field direction. 
The DW propagation speed is proportional to the energy 
dissipation rate that is determined by the DW structure. 
An oscillatory DW motion, either the precession around the 
wire axis or the breath of DW width, should lead to the 
speed oscillation. 
\end{abstract}
\maketitle

Magnetic domain-wall (DW) propagation in a nanowire due to 
a magnetic field\cite{Ono,Cowburn,Erskine,Parkin1,Erskine1}  
reveals many interesting behaviors of magnetization dynamics. 
For a tail-to-tail (TT) DW or a head-to-head (HH) DW (shown 
in Fig. 1) in a nanowire with its easy-axis along the wire 
axis, the DW will propagate in the wire under an 
external magnetic field parallel to the wire axis.
The propagation speed $v$ of the DW depends on the field 
strength\cite{Erskine,Parkin1}. There exists a so-called 
Walker's breakdown field $H_W$\cite{Walker}. $v$ is proportional 
to the external field $H$ for $H<H_W$ and $H\gg H_W$. 
The linear regimes are characterized by the DW mobility $\mu
\equiv v/H$. Experiments showed that $v$ is sensitive to 
both DW structures and wire width\cite{Ono,Cowburn,Erskine}. 
DW velocity $v$ decreases as the field increases between 
the two linear H-dependent regimes, leading to the 
so-called negative differential mobility phenomenon. 
For $H\gg H_W$, the DW velocity, whose time-average is linear 
in $H$, oscillates in fact with time \cite{Walker,Erskine}. 
\begin{figure}[htbp]
 \begin{center}
\includegraphics[width=7.cm, height=4.cm]{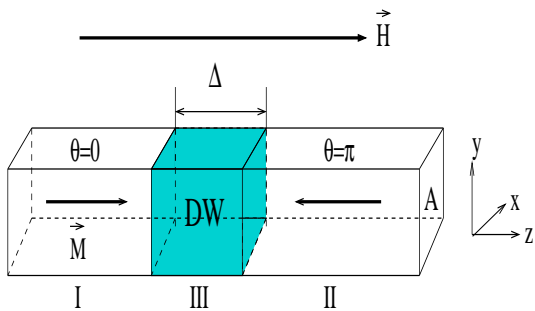}
 \end{center}
\caption{\label{fig1} Schematic diagram of a HH DW of width
$\Delta$ in a magnetic nanowire of cross-section $A$. 
The wire consists of three phases, two domains and one DW. 
The magnetization in domains I and II is along +z-direction 
($\theta=0$) and -z-direction ($\theta=\pi$), respectively. 
III is the DW region whose magnetization structure could 
be very complicate. $\vec H$ is an external field along 
+z-direction.} 
\end{figure}

It has been known for more than fifty years that the 
magnetization dynamics is govern by the Landau-Lifshitz-Gilbert 
(LLG)\cite{gilbert} equation that is nonlinear and can only be 
solved analytically for some special problems\cite{Walker,xrw}. 
The field induced domain-wall (DW) propagation in a strictly 
one-dimensional wire has also been known for more than thirty 
years\cite{Walker}, but its experimental realization in nanowires 
was only achieved\cite{Ono,Cowburn,Erskine,Parkin1,Erskine1} 
in recent years when we are capable of fabricating various nano 
structures. Although much progress\cite{Slon,Thiv} has been made 
in understanding field-induced DW motion, it is still a formidable 
task to evaluate the DW propagation speed in a realistic magnetic 
nanowire even when the DW structure is obtained from various means 
like OOMMF simulator and/or other numerical software packages. 
A global picture about why and how a DW propagates in a magnetic 
nanowire is still lacking. 

In this report, we present a theory that reveals the origin 
of DW propagation. Firstly, we shall show that no static HH 
(TT) DW is allowed in a homogeneous nanowire in the presence 
of an external magnetic field. Secondly, energy conservation 
requires that the dissipated energy must come from the energy 
decrease of the wire. Thus, the origin of DW propagation is 
as follows. A HH (TT) DW must move under an external field 
along the wire. The moving DW must dissipate energy because 
of various damping mechanisms. The energy loss should be 
supplied by the Zeeman energy released from the DW propagation. 
This consideration leads to a general relationship between 
DW propagation speed and the DW structure.  
It is clear that DW speed is proportional to the energy 
dissipation rate, and one needs to find a way to enhance the 
energy dissipation in order to increase the propagation speed. 
Furthermore, the present theory attributes a DW velocity 
oscillation for $H\gg H_W$ to the periodic motion of the DW, 
either the precession of the DW or oscillation of the DW width. 

In a magnetic material, magnetic domains are formed in order to 
minimize the stray field energy. A DW that separates two domains 
is defined by the balance between the exchange energy and the 
magnetic anisotropy energy. The stray field plays little role 
in a DW structure. To describe a HH DW in a magnetic nanowire, 
let us consider a wire with its easy-axis along the wire axis 
(the shape anisotropy dominates other magnetic anisotropies  
and makes the easy-axis along the wire when the wire is small 
enough) which is chosen as the z-axis as illustrated in Fig. 1. 
Since the magnitude of the magnetization $\vec M$ does not 
change in the LLG equation\cite{xrw}, the magnetic state of 
the wire can be conveniently described by the polar angle 
$\theta(\vec x,t)$ (angle between $\vec M$ and the z-axis) and 
the azimuthal angle $\phi(\vec x, t)$. The magnetization energy 
is mainly from the exchange energy and the magnetic anisotropy 
because the stray field energy is negligible in this case. 
The wire energy can be written in general as 
\begin{equation}
\begin{split}
& E=\int F(\theta,\phi,\vec\nabla\theta,\vec\nabla\phi)d^3\vec x,\\ 
& F=f(\theta,\phi)+\frac{J}{2}[(\vec\nabla\theta)^2+\sin^2\theta
(\vec\nabla\phi)^2]-MH\cos\theta, 
\end{split}\label{energy}
\end{equation}
where $f$ is the energy density due to all kinds of magnetic 
anisotropies which has two equal minima at $\theta=0$ and 
$\pi$ ($f(\theta=0,\phi)=f(\theta=\pi,\phi)$), $J-$term is 
the exchange energy, $M$ is the magnitude of magnetization, 
and $H$ is the external magnetic field along z-axis. 
In the absence of $H$, a HH static DW that separates $\theta=0$ 
domain and $\theta=\pi$ domain (Fig. 1) can exist in the wire. 

{\it Non-existence of a static HH (TT) DW in a magnetic 
field}-In order to show that no intrinsic static HH DW is 
allowed in the presence of an external field ($H\neq 0$), 
one only needs to show that following equations have no solution 
with $\theta=0$ at far left and $\theta=\pi$ at far right, 
\begin{equation}
\begin{split}
&\frac{\delta E}{\delta\theta}=J\nabla^2\theta-\frac{\partial f}
{\partial\theta}-HM\sin\theta-J\sin\theta\cos\theta (\vec\nabla\phi)^2=0,\\
&\frac{\delta E} {\delta \phi}=J\vec\nabla\cdot(\sin^2
\theta \vec\nabla\phi)-\frac{\partial f}{\partial \phi}=0. 
\end{split}\label{DW}
\end{equation}
Multiply the first equation by $\nabla\theta$ and the second 
equation by $\nabla\phi$, then add up the two equations. 
One can show a tensor $\hbox{\bf T}$ satisfying 
$\nabla\cdot\hbox{\bf T}=0$ with 
\begin{align}
\hbox{\bf T}=& [f-HM\cos\theta+\frac{J}{2}(|\nabla
\theta|^2+\sin^2\theta|\nabla\phi|^2)]\hbox{\bf 1}-\nonumber\\
& J(\nabla\theta\nabla\theta+\sin^2\theta\nabla\phi\nabla\phi),
\nonumber
\end{align}
where $\hbox{\bf 1}$ is $3\times 3$ unit matrix. A dyadic 
product ($\nabla\theta\nabla\theta$ and $\nabla\phi\nabla\phi$) 
between the gradient vectors is assumed in $\hbox{\bf T}$. 
If a HH DW exists with $\theta=0$ in the far left and $\theta=\pi$ 
in the far right, then it requires $-f(0,\phi)+HM=-f(\pi,\phi)-HM$ 
that holds only for $H=0$ since $f(0,\phi)=f(\pi,\phi)$. 
In other words, a DW in a nanowire under an external field 
must be time dependent that could be either a local motion 
or a propagation along the wire. It should be clear that the 
above argument is only true for a HH DW in a homogeneous wire, 
but not valid with defect pinning that changes Eq. \eqref{DW}. 
Static DWs exist in fact in the presence of a weak field in 
reality because of pinning. 

What is the consequence of the non-existence of a static DW? 
Generally speaking, a physical system under a constant driving 
force will first try a fixed point solution\cite{xrw2}. 
It goes to other types of more complicated solutions 
if a fixed point solution is not possible. 
It means that a DW has to move when an external magnetic field 
is applied to the DW along the nanowire as shown in Fig. 1. 
It is well known\cite{Thiv} that a moving magnetization must 
dissipate its energy to its environments with a rate, 
$\frac{dE}{dt}=\frac{\alpha M}{\gamma}\int_{-\infty}
^{+\infty }\left(d\vec m/dt\right)^2d^3\vec x,$ 
where $\vec m$ is the unit vector of $\vec M$, $\alpha$ and 
$\gamma$ are the Gilbert damping constant and gyromagnetic 
ratio, respectively. Following the similar method in Reference 
12 for a Stoner particle, one can also show that the energy 
dissipation rate of a DW is related to the DW structure as 
\begin{equation}\label{diss}
\frac{dE}{dt}=-\frac{\alpha\gamma}{(1+\alpha^2)M}\int_{-\infty}
^{+\infty }\left(\vec M\times\vec H_{eff}\right)^2d^3\vec x, 
\end{equation}
where $\vec H_{eff}=-\frac{\delta F}{\delta\vec M}$ is the 
effective field. In regions I and II or inside a static DW, 
$\vec M$ is parallel to $\vec H_{eff}$. Thus no energy 
dissipation is possible there. The energy dissipation can 
only occur in the DW region when $\vec M$ is not parallel to 
$\vec H_{eff}$. 

{\it DW propagation and energy dissipation-}For a magnetic 
nanowire in a static magnetic field, the dissipated energy must 
come from the magnetic energy released from the DW propagation. 
The total energy of the wire equals the sum of the energies of 
regions I, II, and III (Fig. 1), $E=E_I+E_{II}+E_{III}$. 
$E_I$ increases while $E_{II}$ decreases when the DW propagates 
from left to the right along the wire. The net energy change 
of region I plus II due to the DW propagation is 
\begin{equation}\label{diss1}
\frac{d(E_{I}+E_{II})}{dt}=-2HMvA,
\end{equation}
where $v$ is the DW propagating speed, and $A$ is the cross 
section of the wire. This is the released Zeeman energy stored 
in the wire. The energy of region III should not change much 
because the DW width $\Delta$ is defined by the balance of 
exchange energy and magnetic anisotropy, and is usually order of 
$10\sim 100nm$. A DW cannot absorb or release too much energy, 
and can at most adjust temporarily energy dissipation rate. 
In other words, $\frac{dE_{III}}{dt}$ is either zero or fluctuates 
between positive and negative values with zero time-average. 
Since energy release from the magnetic wire should be equal to 
the energy dissipated (to the environment), one has 
\begin{equation}
-2HMvA +\frac{dE_{III}}{dt}=-\frac{\alpha\gamma}{(1+\alpha^2)M}
\int_{III}\left(\vec M\times \vec H_{eff}\right)^{2}d^3\vec x. 
\end{equation}
or 
\begin{equation}\label{main}
v= \frac{\alpha\gamma }{2(1+\alpha^2)HA}\int_{III}\left(\vec m\times 
\vec H_{eff}\right)^{2}d^3\vec x +\frac{1}{2HMA}\frac{dE_{III}}{dt}.
\end{equation}

{\it Velocity oscillation-}Eq. \eqref{main} is our central 
result that relates the DW velocity to the DW structure. 
Obviously, the right side of this equation is fully determined 
by the DW structure. A DW can have two possible types of 
motion under an external magnetic field. One is that a DW 
behaves like a {\it rigid body} propagating along the wire. 
This case occurs often at small field, and it is the basic 
assumption in Slonczewski model\cite{Slon} and Walker's 
solution for $H<H_W$. Obviously, both energy-dissipation 
and DW energy is time-independent, $\frac{dE_{III}}{dt}=0$. 
Thus,and the DW velocity should be a constant. The other case 
is that the DW structure varies with time. For example, the 
DW may precess around the wire axis and/or the DW width may 
breathe periodically. One should expect both $\frac{dE_{III}}
{dt}$ and energy dissipation rate oscillate with time. 
According to Eq. \eqref{main}, DW velocity will also oscillate. 
DW velocity should oscillate periodically if only one type of 
DW motion (precession or DW breathing) presents, but it could 
be very irregular if both motions are present and the ratio 
of their periods is irrational. Indeed, this oscillation was 
observed in a recent experiment\cite{Erskine}. How can one 
understand the wire-width dependence of the DW velocity? 
According to Eq. \eqref{main}, the velocity is a functional 
of DW structure which is very sensitive to the wire width. 
For a very narrow wire, only transverse DW is possible while a 
vortex DW is preferred for a wide wire (large than DW width). 
Different vortexes yield different values of $|\vec m\times 
\vec H_{eff}|$, which in turn results in different DW 
propagation speed. 
\begin{figure}[htbp]
 \begin{center}
\includegraphics[width=7.cm, height=4.cm]{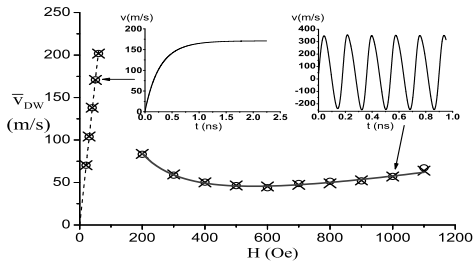}
 \end{center}
\caption{\label{fig2} The time-averaged DW propagation 
speed versus the applied magnetic field for a biaxial 
magnetic nanowire of cross section $4nm\times 20nm$. 
The wire parameters are $K_1=K_2=10^5 J/m^3$, $J=4.\times 
10^{-11} J/m$, $M=10^6 A/m$, and $\alpha=0.1$. Cross are 
for the calculated velocities from Eq. \eqref{ave}, and 
the open circles are for the simulated average velocities. 
The dashed straight line is the fit to the small $H<H_W$ 
results, and solid curve is the fit to $a(H-H_0)^2/H+b/H$. 
Insets: the instantaneous DW speed calculated from Eq. 
\eqref{main} for $H=50Oe <H_W$ (left) and $H=1000Oe>H_W$ 
(right). }
\end{figure}

Time averaged velocity is 
\begin{equation}\label{ave}
\bar v= \frac{\alpha\gamma}{2(1+\alpha^2)HA}\int_{III}
\overline{\left({\vec m\times \vec H_{eff}}\right)^{2}}d^3\vec x, 
\end{equation}
where bar denotes time average. It says that the averaged 
velocity is proportional to the energy dissipation rate. 
In order to show that both Eqs. \eqref{main} and \eqref{ave} 
are useful in evaluating the DW propagation speed from a DW 
structure. We use OOMMF package to find the DW structures 
and then use Eq. \eqref{ave} to obtain the average velocity. 
Figure 2 is the comparison of such calculated velocities 
(cross) and numerical simulation (open circles with their error 
bars smaller than the symbol sizes) for a magnetic nanowire 
of cross-section dimension $4nm\times 20nm$ with a biaxial 
magnetic anisotropy $f=-\frac{K_1}{2}M_z^2+\frac{K_2}{2}M_x^2$. 
The system parameters are $K_1=K_2=10^5 J/m^3$, $J=4.\times 
10^{-11} J/m$, $M=10^6 A/m$, and $\alpha=0.1$. The good overlap 
between the cross and open circles confirm the correctness of 
Eq. \eqref{ave}. The $\bar v -H$ curve for $H>H_W$ can be fit 
well by $a\Delta (H-H_0)^2/H +b/H$ (see discussion later). 
The insets are instantaneous DW propagation velocities for 
both $H<H_W$ and $H>H_W$, by Eq. \eqref{main} from the 
instantaneous DW structures obtained from OOMMF. 
The left inset is the instantaneous DW speed at $H=50Oe<H_W$, 
reaching its steady value in about $1ns$. The right inset is 
the instantaneous DW speed at $H=1000Oe>H_W$, showing clearly 
an oscillation. They confirm that the theory is capable of 
capturing all the features of DW propagation. 

The right side of Eq. \eqref{ave} is positive and non-zero since 
a time dependent DW requires $\vec m\times\vec H_{eff}\neq 0$, 
implying a zero intrinsic critical field for DW propagation.  
If the DW keep its static structure, then the first term in 
the right side of Eq. \eqref{main} shall be proportional to 
$a\Delta AH^2$, where $a$ is a numerical number of order of 
1 that depends on material parameters and the DW structure. 
This is because the effective field due to the exchange energy 
and magnetic anisotropy is parallel to $\vec M$, and does not 
contribute to the energy dissipation. Thus, in this case, 
$v=a\frac{\alpha\gamma \Delta}{1+\alpha^2}H$ with $\mu=a\frac
{\alpha\gamma \Delta}{1+\alpha^2}$. Consider the Walker's 1D 
model\cite{Walker} in which $f=-\frac{K_1}{2}M^2\cos^2\theta+ 
\frac{K_2}{2} M^2 \sin^2\theta\cos^2\phi,$ here $K_1$ 
and $K_2$ describe the easy and hard axes, respectively. 
From Walker's trial function of a DW of width $\Delta$, 
$\ln\tan\frac{\theta(z,t)}{2}=\frac{1}{\Delta(t)}\left[z- 
\int_0^tv(\tau)d\tau \right] $ and $\phi (z,t)=\phi (t),$
one has (from Eq. \eqref{diss}) the energy dissipation rate 
\begin{equation}\label{walker1}
\frac{dE}{dt}=-\frac{2\alpha\gamma A\Delta}{1+\alpha ^{2}}
\left[K_2^2M^{3}\sin^2\phi\cos^2\phi +H^2M\right],  
\end{equation}
and DW energy change rate is 
\begin{equation}\label{walker2}
\frac{dE_{III}}{dt}=\frac{d}{dt}\int_{III}F(\theta,\phi,\vec\nabla
\theta,\vec\nabla\phi)d^3\vec x=-4JA\cdot\frac{\dot\Delta}{\Delta^2}. 
\end{equation}
Substituting Eqs. \eqref{walker1} and \eqref{walker2} into 
Eq. \eqref{main}, one can easily reproduce Walker's DW 
velocity expression for both $H< H_W$ and $\gg H_W$. 
For example, for $H< H_W=\alpha K_2M/2$ and $\Delta=const.$, 
Eq. \eqref{main} gives 
\begin{equation}
v=\frac{\alpha\Delta\gamma}{1+\alpha^2}\left[1+
\left(\frac{K_2M\sin\phi\cos\phi}{H}\right)^2\right]H.
\end{equation}
This velocity expression is the same as that of the 
Slonczewski model\cite{Slon} for a one-dimensional wire. 
In Walker's analysis, $\phi$ is fixed by $K_2$ and $H$ 
through $K_2M\sin\phi\cos\phi=\frac{H}{\alpha }$. 
Using this $\phi$ in the above velocity expression, Walker's 
mobility coefficient $\mu=\frac{\gamma\Delta}{\alpha}$ 
is recovered. This inverse damping relation is from the 
particular potential landscape in $\phi$-direction. 
One should expect different result if the shape of the 
potential landscape is changed. Thus, this expression should 
not be used to extract the damping constant\cite{Ono,Erskine}.

A DW may precess around the wire axis as well as be 
substantially distorted from its static structure when 
$H>H_W$ as it was revealed in Walker's analysis. According 
to the minimum energy dissipation principle\cite{Mea}, a DW 
will arrange itself as much as possible to satisfy Eq. (2). 
Thus, the distortion is expected to absorb part of $H$. 
The precession motion shall induce an effective field 
$g(\phi)$ in the transverse direction, where $g$ depends 
on the magnetic anisotropy in the transverse direction. 
One may expect $\vec m\times\vec H_{eff}\simeq (H-H_0) 
\sin\theta\hat{z}+\sin\theta g(\phi)\hat{y}$, where 
$H_0$ is the DW distortion absorbed part of $H$.  
Using $|\vec m\times\vec H_{eff}|^2=(H-H_0)^2\sin^2\theta 
+g^2\sin^2\theta$ in Eq. \eqref{ave}, the DW propagating 
speed takes the following h-dependence, $v=a\alpha\gamma
\Delta (H-H_0)^2/H(1+\alpha^2)+b\alpha\gamma\Delta /[H(1+
\alpha^2)]$, linear in both $\Delta$ and $H$ for $H\gg H_0$, 
but a smaller DW mobility. This field-dependence is supported 
by the excellent fit in Fig. 2 for $H>H_W$. The reasoning agrees 
also with the minimum energy dissipation principle\cite{Mea} 
since $|\vec m\times H_{eff}|=H\sin\theta$ when $\vec M$ for 
$H=0$ is used, and any modification of $\vec M$ should only 
make $|\vec m \times H_{eff}|$ smaller. The smaller mobility 
at $H\gg H_W, H_0$ leads naturally to a negative differential 
mobility between $H<H_W$ and $H\gg H_W$! In other words, the 
negative differential mobility is due to the transition of the 
DW from a high energy dissipation structure to a lower one. 
This picture tells us that one should try to make a DW 
capable of dissipating as much energy as possible if one 
wants to achieve a high DW velocity. This is very different 
from what people would believe from Walker's special mobility 
formula of inverse proportion of the damping constant. 
To increase the energy dissipation, one may try to reduce 
defects and surface roughness. 
The reason is, by minimum energy dissipation principle, that 
defects are extra freedoms to lower $|\vec m\times \vec H_{eff}|$ 
because, in the worst case, defects will not change $|\vec 
m\times \vec H_{eff}|$ when $\vec M$ without defects are used. 

The correctness of our central result Eq. \eqref{main} depends 
only on the LLG equation, the general energy expression of Eq. 
\eqref{energy}, and the fact that a static magnetic field can 
be neither an energy source nor an energy sink of a system. 
It does not depend on the details of a DW structure as long 
as the DW propagation is induced by a static magnetic field. 
In this sense, our result is very general and robust, and  
it is applicable to an arbitrary magnetic wire. 
However, it cannot be applied to a time-dependent field or 
the current-induced DW propagation, at least not directly.  
Also, it may be interesting to emphasize that there is no 
inertial in the DW motion within LLG description since this 
equation contains only the first order time derivative. 
Thus, there is no concept of mass in this formulation. 

In conclusion, a global view of the field-induced DW propagation 
is provided, and the importance of energy dissipation in the 
DW propagation is revealed. A general relationship between the 
DW velocity and the DW structure is obtained. The result says: 
no damping, no DW propagation along a magnetic wire. 
It is shown that the intrinsic critical field for a HH 
DW is zero. This zero intrinsic critical field is related to 
the absence of a static HH or a TT DW in a magnetic field 
parallel to the nanowire. Thus, a non-zero critical field can 
only come from the pinning of defects or surface roughness.  
The observed negative differential mobility is due to 
the transition of a DW from a high energy dissipation 
structure to a low energy dissipation structure. Furthermore, 
the DW velocity oscillation is attributed to either the DW 
precession around wire axis or from the DW width oscillation. 

This work is supported by Hong Kong UGC/CERG grants 
(\# 603007 and SBI07/08.SC09).

\end{document}